\documentclass[amsmath,amssymb,aps,prl,twocolumn,groupedaddress,superscriptaddress,10pt,reprint,longbibliography]{revtex4-1}
\usepackage{graphicx}
\usepackage{textcomp} 
\usepackage{amsmath}
\usepackage{color}
\usepackage[colorlinks,linkcolor=blue,citecolor=blue,urlcolor=blue]{hyperref}

\begin{document}

\title{Polarization Entanglement from Parametric Down-Conversion with a LED Pump }

\author{Wuhong Zhang}
\email{zhangwh@xmu.edu.cn}
\author{Diefei Xu}
\author{Lixiang Chen}
\email{chenlx@xmu.edu.cn}
\affiliation{Department of Physics, Xiamen University, Xiamen 361005, China}

	\date{\today}
\begin{abstract}
Spontaneous parametric down-conversion (SPDC) is a reliable platform for entanglement generation. Routinely, a coherent laser beam is an essential prerequisite for pumping the nonlinear crystal. Here we break this barrier to generate polarization entangled photon pairs by using a commercial light-emitting diode (LED) source to serve as the pump beam. This effect is counterintuitive, as the LED source is of extremely low spatial coherence, which is transferred during the down-conversion process to the biphoton wavefunction. However, the type-II phase-matching condition naturally filters the specific frequency and wavelength of LED light exclusively to participate in SPDC such that localized polarization Bell states can be generated, regardless of the global incoherence over the full transverse plane. In our experiment, we characterize the degree of LED light-induced polarization entanglement in the standard framework of the violation of Bell inequality. We have achieved the Bell value $S=2.33\pm 0.097$, obviously surpassing the classical bound $S=2$ and thus witnessing the quantum entanglement. Our work can be extended to prepare polarization entanglement by using other natural light sources, such as sunlight and bio-light, which holds promise for electricity-free quantum communications in outer space.

\end{abstract}

\maketitle

Quantum entanglement is generally regarded as a vital characteristic of quantum information theory \cite{Bennet1995Phystoday,Nielsen2002quantum}. Meanwhile, photons are one of the best physical systems for the long-distance transmission of quantum states. Currently, the most mature method of generating photonic entanglement is spontaneous parametric down-conversion (SPDC), where a pump photon is converted into two photons of lower frequencies under the conservation of energy and momentum in a nonlinear crystal \cite{mandel1995Book}. This  process of generating photons in pairs, makes it easy to entangle two photons in various degrees of freedom. On account of the extraordinary precision, polarization-entangled photon pairs are easy to generate, control, and measure, and have become a major tool for quantum teleportation \cite{bouwmeester1997nature}, dense encoding \cite{Mattle1996PhysRevLett}, quantum cryptography \cite{Jennewein2000PhysRevLett}, quantum holography \cite{defienne2021polarization}, quantum computing \cite{o2003demonstration,o2007optical}, quantum information with integrated devices \cite{sansoni2014integrated}, long-distance quantum communication \cite{Ursin2007natphys,ma2012nature}, and even in quantum satellite communication schemes \cite{Yin2017PhysRevLett}.

In the framework of SPDC, polarization-entangled photon pairs have been produced with one crystal in non-collinear \cite{kwiat1995new,increased2004lee} and collinear \cite{collinear2008trojek} phase matching, two-crystal geometries \cite{kwiat1999ultrabright}, Sagnac interferometers \cite{generation2004shi}, nested interferometers \cite{fiorentino2004generation}, time-reversed Hong-Ou-Mandel interference \cite{polarization2018chen}, femtosecond-pulse-pumped lasers \cite{generation2002nambu}, and even in the fiber communication band \cite{bright2010evans}. However, all of the above works utilize a spatially coherent laser as the pump so that the laser appear to be necessary for the production of entangled photon pairs with SPDC.

On the other hand, the influences of the spatial coherence of the pump on the down-converted photon pairs have recently reached great interest. There are some early theoretical studies of the transfer of the pump's spatial coherence \cite{Jha10,Giese18,vintskevich2019effect,Bhaskar2020josab,joshi2020spatial} and the degree of polarization \cite{Kulkarni2016PhysRevA,patoary2019intrinsic,sharma2021controlling} to the down-converted photon pairs. By using a diffuser to transform the laser into a partially spatial coherent pump beam, the generation of spatial \cite{defienne2019spatially} and polarization \cite{ismail2017polarization} entangled photon pairs have been implemented experimentally. Interestingly, pumping by a twisted Gaussian Schell-model beam, the effect of the amount of spatial  entanglement increases inversely with the degree of coherence, which has been demonstrated theoretically \cite{hutter2020boosting}. Even more, a commercially light-emitting diode (LED), as a totally incoherent pump beam, has been used to induce the SPDC process and to examine correlated photon pairs \cite{tamovsauskas2010observation,galinis2011photon,galinis2012modeling,nishii2019heralded}. The influence of the pump coherence on the generation of position-momentum entanglement has been observed experimentally \cite{zhang2019influence}. The entangled photon pairs produced from spatial incoherent pump beams have been proven to be more robust to the effects arising from atmospheric turbulences, which might hold some advantages for free-space quantum information \cite{Qiu2012applphysb,Gbur2014joptsocama,Phehlukwayo2020PhysRevA}. Even though previous studies \cite{sharma2021controlling,ismail2017polarization} have shown that it is in principle able to obtain polarization-entangled photon pairs from the spatial incoherent pump, an experimental technique realization, especially from a truly incoherent source (such as a commercial LED), has not yet been studied very well.

\begin{figure*}[t]
\centerline{\includegraphics[width=1.9\columnwidth]{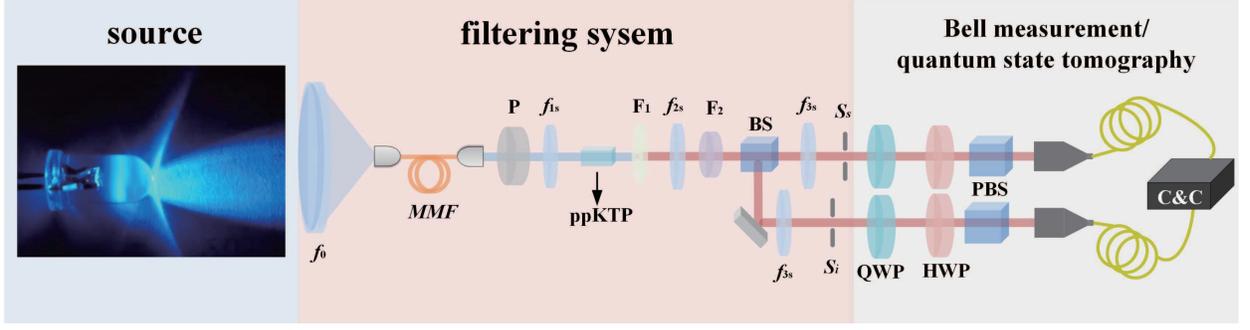}}
\caption{Schematic illustration of the experimental setup, see text for more details.}
\label{fig1}
\end{figure*}

In this work, we pump a nonlinear crystal with a truly spatiotemporally incoherent beam, a commercial LED, to produce polarization entangled photon pairs. We strictly demonstrate the polarization entanglement by observing a violation of a Clauser-Horne-Shimony-Holt (CHSH) inequality \cite{Clauser1969PhysRevLett} both in theory and experiment. Moreover, we quantitatively analyze the quantum state of the down-converted photon pairs by exploiting quantum state tomography under different collection setups. The non-local correlation of the produced photon pairs is verified by a bell value of about 2.3 as well as a concurrence of about 0.63. We observe a good agreement between experiment and theory when it comes to the degree of entanglement of the bi-photon states after the filtering system. Our technique to obtain polarization-entangled photon pairs from a cheap LED instead of an expensive laser may promote a further application, such as producing polarization entangled photon pairs even from sunlight.

 Our experimental setup is shown in Fig.\ref{fig1}, which mainly contains three parts: the incoherent source, the filtering system and the measurement part. The incoherent source is a commercial LED (Thorlabs M405L3) centered at 405 nm in the blue spectral range. The band width of the LED is about 20 nm and the maximum intensity of the LED is 980 mW. Then the followed part is a filtering system that can be used to produce polarization entangled photons. To ensure a Gaussian-like beam profile while maintaining transverse incoherence, we use a collimating lens (Thorlabs COP1-A) followed by a short focal length ($f_0=16$ mm) to couple the incoherent source into a 400-\textmu m-core multimode fiber (MMF). After passing a polarizer (P), the out-coupled LED beam is demagnified by a 4$f$-system ($f_{1s}$:$L_1=300$ mm, $L_2=75$ mm) to have a proper pump beam size ($\sigma_0=0.8$ mm ) when it enters the crystal. The pump power before the crystal is 290 \textmu W, which is just about 3\textperthousand of the LED source. The key part of the filtering system is the nonlinear crystal, a 1\,mm$\times$2\,mm$\times$2.5\,mm periodically-poled potassium titanyl phosphate crystal (ppKTP), which is phase-matched for type-II collinear emission. It is noted that the length of the ppKTP is 2.5 mm to ensure a maximal number of photon counts while maintaining minimal walk-off effects between the two polarizations of signal and idler so that we can omit a timing compensation crystal \cite{kuklewicz2004high,sharma2021controlling}.
 \begin{figure*}[t]
\centerline{\includegraphics[width=1.8\columnwidth]{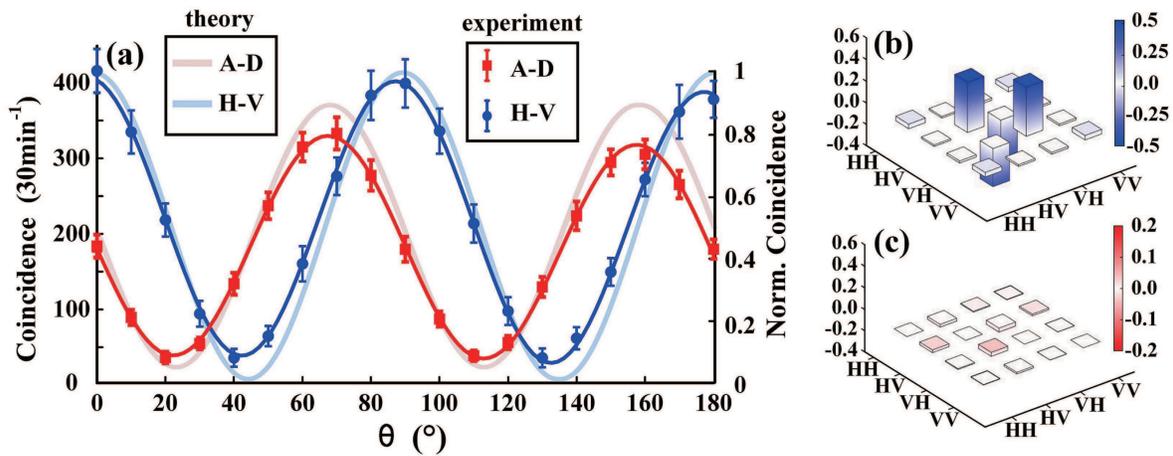}}
\caption{(a) Theoretical plot and experimental results of the coincidence photon count in A-D and H-V basis. The red and blue lines correspond to the $\sin^2-fit$ function of the averaged experimental coincidence counts, while the light red and light blue lines correspond to the theoretical calculation. (b) and (c) are the real and imaginary part of the measured density matrix. }
\label{fig2}
\end{figure*}
A long-pass filter ($F_1$) and a 3 $nm$ spectral filter at 810 nm ($F_2$) after the crystal are used to block the pump beam and to ensure that only degenerate photons are measured. Then another 4$f$-system ($f_{2s}$: $L_1=50$ mm, $L_2=100$ mm) is used to image the crystal onto the face of a 50:50 non-polarizing beam splitter (BS). Both photons of each pair are equally to be transmitted or reflected at the BS. Hence, on average half of the generated pairs yield one photon in the transmitted path and one in the reflected path. In these incidences two-photon states produced by SPDC can be represented by a density matrix of the form:
\begin{equation}
\begin{aligned}
&\hat{\rho}=|A|^2\int{d\boldsymbol{q}_sd\boldsymbol{q}_id\boldsymbol{q}_{s}^{'}d\boldsymbol{q}_{i}^{'}}\langle G( \boldsymbol{q}_s+\boldsymbol{q}_i) G^*( \boldsymbol{q}_{s}^{'}+\boldsymbol{q}_{i}^{'})\rangle \\
&\times\operatorname{sinc}^2( \frac{\varDelta qL}{2}) \frac{1}{\sqrt{2}}(|H_s,\boldsymbol{q}_s\rangle|V_i,\boldsymbol{q}_i\rangle +|V_s,\boldsymbol{q}_s\rangle|H_i,\boldsymbol{q}_i\rangle)
\\
&\times \frac{1}{\sqrt{2}}({\langle H_{s'},\boldsymbol{q}_{s}^{'}|\langle V_{i'},\boldsymbol{q}_{i}^{'}|+\langle V_{s'},\boldsymbol{q}_{s}^{'}|\langle H_{i'},\boldsymbol{q}_{i}^{'}|})
\label{Eq1}
\end{aligned}
\end{equation}
where $|A|^2$ is a constant that depends on physical constants, $\langle G(\cdot)G^*(\cdot)\rangle$ represents the angular correlation function of the pump field, $ \boldsymbol{q}_s,\boldsymbol{q}_s^{'} $ and $\boldsymbol{q}_i,\boldsymbol{q}_i^{'}$ are transverse wave vectors of the signal and idler fields.  $\operatorname{sinc}^2(\varDelta qL/2)$ denotes the phase matching function of the nonlinear crystal. The index $s, i$ denotes the spatial mode called signal, idler. H and V are the polarization state of the photon.  $\left|H_s,\boldsymbol{q}_s\right>$ and $\left|V_i,\boldsymbol{q}_i\right>$ are the states for the mode with polarization and transverse wave vector of the signal and idler. The other degree of freedom, such as the spatial freedom of the photon in this mode is encoded in its transverese wave vector.
Based on Eq. \ref{Eq1}, the state is determined by a mode function which is associated with the pump angular spectrum as well as the phase matching function of the crystal, and the polarization projection.
 If one photon is detected at position $ \boldsymbol{r}_s$ at time $t$ with the polarization angle $\theta_s$ while another photon is detected at position $ \boldsymbol{r}_i$ at time $t+\tau$ with the polarization angle $\theta_i$, the coincidence count rate of the bi-photon states can be written as:
\begin{equation}
\begin{aligned}
&R_{si}(\theta _s,\theta _i ,\boldsymbol{r}_s,\boldsymbol{r}_i) =\Bigg|| A|^2e^{ik_p(\boldsymbol{r}_s+\boldsymbol{r}_i )}\\
&\times\Bigg[\frac{A_p\pi \ell_c{\sigma _0}^2}{z\sqrt{4{\sigma _0}^2+\ell _{c}^{2}}}\Bigg]^2 \Bigg[C_1C_2\cdot \frac{\pi z^2}{2k_{p}^{2}\sigma _z\delta}\sin 2\theta _s\sin 2\theta _i\\
&+\sqrt{\frac{2\pi}{k_{p}^{2}\delta ^2}}(C_{1}^{2}\cos ^2\theta _s\sin ^2\theta_i+C_{2}^{2}\sin ^2\theta _s\cos ^2\theta_i) \Bigg]\Bigg|,
\label{EqB}
\end{aligned}
\end{equation}
where $k_p$ is the pump wave vector, $\sigma _0 $ is the beam size, while the $\sigma _{z}$ is the beam size at position $z$ along the propagating direction, which satisfies the relationship of $\sigma _z={{z\sqrt{\ell _{c}^{2}+4\sigma _{0}^{2}}}/{2k_p\sigma _0\ell _c}}$, and $ \ell_c $ is the transverse coherence length, $\delta =2\ell _c\sigma _z/\sqrt{4{\sigma _z}^2+\ell _{c}^{2}}$ is the effective spectral width \cite{mandel1995Book}, $C_1, C_2$ denote the detection efficiency of the whole system, $A_p$ is a constant induced by the cross-spectral density function of a Gaussian Schell Model (GSM) pump beam. The detailed derivation of Eq. (\ref{EqB}) can be found in [Supplemental Material]. Based on Eq. (\ref{EqB}), one can see that the count rate is independent of the position but just up to a phase factor, while the beam size $\sigma_z$ and the spatial coherence $\delta$ will influence the count rate. Considering the Gaussian profile of the pump beam, to get the highest count rate of the position correlation of signal and idler photons, we need to scan the detectors position firstly.

\begin{figure*}[t]
\centerline{\includegraphics[width=1.8\columnwidth]{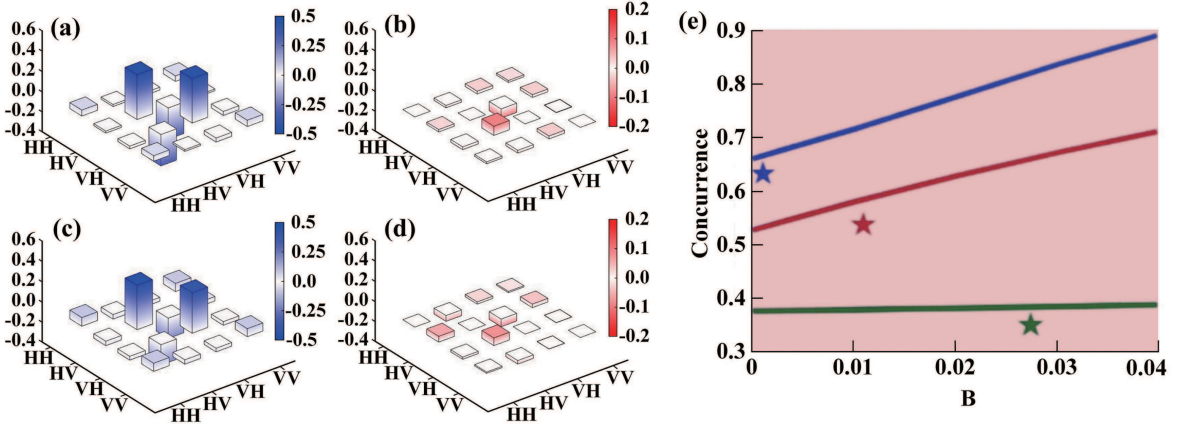}}
\caption{(a)-(d)Quantum state tomography of the output states collected by two different multi-mode fibers (MMF). (a) and (b) represent the real and the imaginary parts of the state with a $50\mu m$ MMF. (c) and (d) represent the real and the imaginary parts of the state with a $125\mu m$ MMF. (e) the relationship of the concurrence with the spatial coherence of pump in simulation (lines) and the experimental results (stars).}
\label{fig3}
\end{figure*}

 To increase the detectable signal, instead of scanning the point detector, our proposed technique is to scan the small range of the signal and idler photons to find the initial position correlated photons, which is based on our recent work that uses an LED to implement the position correlation detection of signal and idler photons \cite{zhang2019influence}.
 In order to simplify our measurement setup, we restrict the transverse position $\boldsymbol{\rho }_s,\boldsymbol{\rho }_i$ to one direction denoted by $y_s, y_i$. Therefore, we re-image ($f_{3s}: L_1=L_2=100$ mm) the facet of the BS onto the facet of the two single slit ($S_s, S_i$). The width of each slit is set about $0.1$ mm to guarantee the quality of the photon pairs' position correlation in $y$ direction [see Supplemental Material for more details].

The polarization projection angles of signal and idler are performed by a half wave plate (HWP) together with a polarizing beam splitter (PBS) in each arm, as shown after the filtering system in Fig. \ref{fig1}. By rotating the angle of HWP, we can realize a Bell measurement of the bi-photons. After that, a microscope objectives collects the photons and then couples them into a SMF connected with the single photon detector. The distance from the slit to the single photon detector is about $z=100$ mm that is also used in our theoretical analysis with Eq. (\ref{EqB}).
The photon coincidence count rate is recorded with a coincidence window of 1 ns. Since the incoherent pump power is only about 290 \textmu W at the plane of the crystal, we observe a maximum coincidence count rate of about 15 counts/min, while the maximum single photon count rate is about 10000 counts/min.

Our experimental results are shown in Fig. \ref{fig2}. The plots of coincidence counts for A-D (red rectangular points) and H-V (blue points) as a function of the rotation angle ($\theta$) of the HWP are shown in Fig. \ref{fig2}(a). Accidental coincidences have been removed in these plots and the sinusoidal fits  (red solid line for A-D and blue solid line for H-V ) are used to obtain the respective visibility.  All of the data points are accumulated by 60 measurements. Each measurement is measured in 30s and averaged 60 times to get the error bar. The visibility of H-V basis is about 86$\%$ while the A-D basis is about 82$\%$. Based on the CHSH form of Bell's inequality \cite{Clauser1969PhysRevLett},  we use the 16 combinations of the HWP settings ($\theta_1=0^{\circ}, 22.5^{\circ}, 45^{\circ}, 67.5^{\circ}; \theta_2=11.25^{\circ}, 33.75^{\circ}, 56.25^{\circ}, 78.75^{\circ}$) to calculate the Bell parameter $S=2.33\pm 0.097$ experimentally. It is noted that the theoretical value of the coincidence count rate can be also plotted based on Eq.(\ref{EqB}), as shown in the Fig. \ref{fig2}(a) (light red solid line for A-D and light blue solid line for H-V ). The theoretical Bell parameter is $S=2.61$. Here, the transverse coherent length $\ell_c$ is estimated to be 11 $\mu m$ based on our previous work \cite{zhang2019influence}, and the beam waist is measured several times and averaged approximately as $\sigma_0=5$ mm. In our measurement setup, the slit as well as the detection system is not a strictly point detector as the theoretical assumption. Therefore, the diameter of the collection setup determines the number of the collected modes and thus lowers the value of Bell parameter experimentally.

Apparently, the produced polarization entangled state is not a maximally entangled state.
In order to better illustrate the polarization entangled photon pairs extracted from an LED pump, we show a more detailed characterization of the output state obtained by quantum state tomography. For that, a quarter-wave plates (QWP) is inserted before the HWP in each arm of the light path. Then we can quantitatively analyze the concrete quantum state of the detected photon pairs. The density matrix $\hat{\rho}$ of the output state can be reconstructed through a series of projection measurements. The individual density matrix elements can be obtained from the coincidence counts recorded in the corresponding measurements by a maximum likelihood algorithm. We estimate $\hat{\rho}$ from the 16 projection measurements by using the methods outlined in Ref \cite{James2001PhysRevA}. The real and imaginary parts of $\hat{\rho}$ are shown in Fig. \ref{fig2} (b) and (c), respectively. Based on the measured density matrix $\hat{\rho}$, we can also obtain that the concurrence, which is used to represent the degree of entanglement \cite{PhysRevLett.Concurrence}, is about 0.63. 

 It is noted that previous works have shown that the multimode nature of the down-converted photon pairs when using a LED as the pump source \cite{galinis2011photon,galinis2012modeling}. 
 Besides, the number of the spatial modes increases with increasing the beam size $\sigma_z$ of the LED \cite{tamovsauskas2010observation}. To further verify the influence of the purity of the collected modes on concurrence for LED pump,
 we replace the SMF (core diameters $5\mu m$) with two kinds of multi-mode fiber (MMF) which the core diameters are $50\mu m$ and $125\mu m$ respectively. Increasing the core diameters of the collection fiber means we have collected more spatial modes than just the Gaussian modes by SMF. Similarly, the quantum state tomography is conducted to obtain the corresponding density matrix of the collected two-photon state as shown in Fig. \ref{fig3}. Based on these density matrixs, one can calculate the concurrence, which is about 0.54 for $50\mu m$ MMF and 0.35 for $125\mu m$ MMF, respectively. The experiment result reveals that the concurrence is lower in the case of the multi-mode fiber collection, and decreasing with the increasing of the fiber cross-sectional diameter. In fact, compared with coupling the single-element detector through a multi-mode fiber, the single mode one is considered as a way for mode selectivity. What's more, the spatial selectivity means that the signal photon is similarly conditioned to the idler one, as a result, the polarization entanglement is of high quality.

Based on Eq. (\ref{Eq1}) and Eq.(\ref{EqB}), we can actually predict the influence of the beam size $\sigma_z$ and the spatial coherence $\delta$ on the value of concurrence.
 we scale the collection diameter of the fiber proportional to the beam size, which $5mm$ is for SMF, $0.5mm$ is for $50\mu m$ MMF and $0.2mm$ is for $125\mu m$ MMF, such that the variation of the value of concurrence with spatial coherence under the different $\sigma_z$ can be plotted, as the blue (upper), red (center), and green(lower) line illustrated in Fig. \ref{fig3}(e). It is noted that the B of x-axis is expressed as $B=\delta/{2\sigma_z}$ adopted to quantify the degree of beam's spatial coherence, which interlinks the beam size and transverse correlation length $\ell_c$. Then
 we can plot our experimental points with the value of concurrence and B, as the blue star (upper), red star (center), and green star(lower) shown in \ref{fig3}(e). One can see that the experimental points are slightly lower than the theoretical predictions due to the imperfect experimental conditions. And we expect the value can be improved with smaller collection setups as the theoretical blue line (upper) predicted. In this regard, our proposed method also provides a versatile polarization entangled source that can tune the degree of entanglement as demanded. We also noticed that our experimental results demonstrate the theoretical prediction discussed in a recent related topic work \cite{li2022experimental}.

In summary, we have successfully produced polarization-entangled photon pairs from a truly spatiotemporally incoherent pump beam, a commercial LED. The theoretical derivation of the influence of the pump transverse coherence length on the count rate of bi-photons in SPDC under the post-selection process has been well performed. We have experimentally achieved the violation of Bell inequality with $S=2.33\pm 0.097$, and thus witnessing the quantum entanglement definitely. The experimental results are well consistent with our theoretical prediction. Our technique, obtaining polarization entangled photon pairs from a cheap LED instead of a laser, may promote the further application, such as producing the polarization entangled photon pairs even from sunlight, which provides a way to obtain the electricity-free polarization entangled photon source for outer space quantum communication.

\begin{acknowledgments}
We are grateful to R. W. Boyd, R. Fickler and E. Giese for discussions initiating this project.
Moreover, we thank in particular E. Giese for his support of the analysis of the theory discussions and thank B. Braverman, G. Kulkarni, C. Li as well as the Quantum Nonlinear Optics and Quantum Photonics at the University of Ottawa for fruitful discussions.
This work is supported by the National Natural Science Foundation of China (12034016, 61975169 and 11904303), the Fundamental Research Funds for the Central Universities at Xiamen University (20720220030, 20720200074), the Natural Science Foundation of Fujian Province of China for Distinguished Young Scientists (2015J06002), the Youth Innovation Fund of Xiamen (3502Z20206045).\\
W. Zhang and D. Xu contributed equally to this work.
\end{acknowledgments}

%

\end{document}


\title{Supplemental material: Polarization Entanglement from Parametric Down-Conversion with a LED Pump }

\author{Wuhong Zhang}
\email{zhangwh@xmu.edu.cn}
\author{DieFei Xu}
\author{Lixiang Chen}
\email{chenlx@xmu.edu.cn}
\affiliation{Department of Physics, Xiamen University, Xiamen 361005, China}
	\date{\today}
\maketitle

Generally, the down-converted photon pairs are generated through the SPDC process, which has been well formulated with the Hamiltonian of a quantized pump field, and see more texts in the Appendix A and B of Ref \cite{Giese18}. As the nonlinear crystal we used in our experiment, which is cut for type II phase matching, the photon pairs are indistinguishable in frequency, which is quite useful to entangle them in polarization. In our experiment, we have adopted a beam splitter (BS) to generate the desired polarization-entangled state in 50$\%$ of the time. Specifically, if the down-converted photons are fed into a port of a 50-50 BS, each of the two photons, distinguishable in polarization, has the equal chance of being reflected or transmitted. At the either output port of the BS, the detector is used to monitor the desired coincidence in this way. One can discards the cases in which both photons of a pair exit from the same BS output port, which is called post-selection. Because of the narrow bandwidth of the filters, one can assume that each of the detectors picks up only the monochromatic signal and idler field. Considering the spatial incoherent pump beam together with the mentioned configuration, the two-photon state after the BS can be represented by a density matrix which can be written as:

\begin{equation}
\begin{aligned}
\rho &=| A |^2\int{d\boldsymbol{q}_sd\boldsymbol{q}_id\boldsymbol{q}_{s}^{'}d\boldsymbol{q}_{i}^{'}}< G( \boldsymbol{q}_s+\boldsymbol{q}_i) G^*( \boldsymbol{q}_{s}^{'}+\boldsymbol{q}_{i}^{'})> \\
&\times \operatorname{sinc}^2( \frac{\varDelta qL}{2}) \frac{1}{\sqrt{2}}(|H_s,\boldsymbol{q}_s>\left| V_i,\boldsymbol{q}_i \right> +|V_s,\boldsymbol{q}_s>|H_i,\boldsymbol{q}_i>)
\\
&\times \frac{1}{\sqrt{2}}(<H_{s'}\boldsymbol{q}_{s}^{'}|<V_{i'}\boldsymbol{q}_{i}^{'}|+<V_{s'}\boldsymbol{q}_{s}^{'}|<V_{i'}\boldsymbol{q}_{i}^{'}|)
\label{EqA1}
\end{aligned}
\end{equation}
where $|A|^2 $ is a constant that depends on physical constants, $ < \cdots > $ represents the ensemble average over the different realizations of the pump while being recognized as the angular correlation function of the pump field, $ \boldsymbol{q}_s $ and $\boldsymbol{q}_i$ are transverse wave vectors of the signal and idler fields.

Consider the fact that the transverse dimensions of the crystal are much larger than the pump wavelength, the significant contributions of the $\operatorname{sinc}$ functions to the integral when $ \varDelta k_x<{{\pi k_p}\big/{2\varOmega}} $, where $ \varOmega $ is the ratio of the crystal $ L_x $ to the pump wavelength $ \lambda _p $. As the crystal is much wider than a pump wavelength, $ \varOmega $ is large while the $ \operatorname{sinc}$function only contributes significantly when $ \varDelta k_x $ is only a very small fraction of $ k_p $, thus the $\operatorname{sinc}$ function acts like a delta function in this condition, setting $ \vec{q}_p=\vec{q}_s+\vec{q}_i$.

Furthermore, the coincidence count rate $R_{si}( \theta_s,\theta_i,\boldsymbol{r}_s,\boldsymbol{r}_i)$, the probability per $(unit\,\,time)^2 $, which a photon is detected at position $\boldsymbol{r}_s$ at time t with the polarization angle $\theta_s$ of the polarizer inserted in the path of the signal field while another at position $\boldsymbol{r}_i$ at time $t+\tau $ with the polarization angle $\theta_i$ of the polarizer inserted in the path of the idle field, which can be written as $R_{si}(\theta_s,\theta _i,\boldsymbol{r}_s,\boldsymbol{r}_i)=\alpha _s \alpha _iTr [\rho\hat{E}_{s}^{(-)}(\theta _s,\boldsymbol{r}_s,t)\hat{E}_{i}^{(-)}(\theta _i,\boldsymbol{r}_i,t+\tau) \hat{E}_{i}^{(+)}(\theta_i,\boldsymbol{r}_i,t+\tau ) \hat{E}_{s}^{( + )}(\theta_s,\boldsymbol{r}_s,t)]$ \cite{PhysRev1963Glauber}. Here, $\alpha_s$ and $\alpha_i$ denote the quantum efficiencies of the detectors $D_s$ and $D_i$ located at $ \boldsymbol{r}_s$ and $\boldsymbol{r}_i$, respectively, where the indicator $ \boldsymbol{r}_j$ is defined as $\boldsymbol{r}_j\equiv (\boldsymbol{\rho}_j,z ) $, and $j=i, s$. Meanwhile, the positive frequency parts of the electric fields at detectors $ D_s $ and $ D_i $ are denoted as $ \hat{E}_{s}^{(+)}$ and $\hat{E}_{i}^{(+)}$, respectively, which are equal to the sum of the signal and idler fields arriving at detectors $D_s$ and $D_i$ in the two alternatives. And the two alternatives are caused by the possibility that the down-converted photons can pass or reflect at the BS and get detected in coincidence, with the specific expressions

\begin{equation}
\begin{aligned}
\hat{E}_{i}^{( + )}( \theta _i,\boldsymbol{r}_i,t ) &=b_{i1}\hat{E}_{i1}^{( + )}( \theta _i,\boldsymbol{r}_{i1} ) e^{-i\omega _i( t-t_{i1} )}\\
&+b_{i2}\hat{E}_{i2}^{( + )}( \theta _i,\boldsymbol{r}_{i2} ) e^{-i\omega _i( t-t_{i2} )}
\\
\hat{E}_{s}^{( + )}( \theta _s,\boldsymbol{r}_s,t ) &=b_{s1}\hat{E}_{s1}^{( + )}( \theta _s,\boldsymbol{r}_{s1} ) e^{-i\omega _s( t-t_{s1} )}\\
&+b_{s2}\hat{E}_{s2}^{( + )}( \theta _s,\boldsymbol{r}_{s2} ) e^{-i\omega _s( t-t_{s2} )}.
\label{EqA2}
\end{aligned}
\end{equation}
The constant factors $b_{i1}$ and $b_{s1}$ depend on the transmitted and reflected coefficient of the BS.
After substituting Eq. (\ref{EqA2}), one can rewrite the coincidence count rate $ R_{si}( \theta _s,\theta _i,\boldsymbol{r}_s,\boldsymbol{r}_i ) $, which is also known as the interference law for the two-photon field with the formulation as follows:

\begin{equation}
\begin{aligned}
R_{si}( \theta _s,\theta _i,\boldsymbol{r}_s,&\boldsymbol{r}_i ) =C_1C_2e^{i[ \omega _s( t_{s1}-t_{s2} ) +\omega _i( t_{i1}-t_{i2} ) ]}
\\
&\times W^{( 2 )}( \boldsymbol{\rho }_{s1},\boldsymbol{\rho }_{i1},\boldsymbol{\rho }_{s2},\boldsymbol{\rho }_{i2},\theta _s,\theta _i,z )
\\
&+C_{1}^{2}S^{( 2 )}( \boldsymbol{\rho }_{s1},\boldsymbol{\rho }_{i1},\theta _s,\theta _i,z )
\\
&+C_{2}^{2}S^{( 2 )}( \boldsymbol{\rho }_{s2},\boldsymbol{\rho }_{i2},\theta _s,\theta _i,z ) +c.c.
\label{EqA3}
\end{aligned}
\end{equation}
where $ C_1=\sqrt{\alpha _s\alpha _i}b_{s1}b_{i1} $, $ C_2=\sqrt{\alpha _s\alpha _i}b_{s2}b_{i2} $. What's more, the term $ W^{( 2 )}( \boldsymbol{\rho }_{s1},\boldsymbol{\rho }_{i1},\boldsymbol{\rho }_{s2},\boldsymbol{\rho }_{i2},\theta _s,\theta _i,z ) $ in the Eq. (\ref{EqA3}) is usually regarded as the two-photon cross-spectral density function with the definition:
\begin{equation}
\begin{aligned}
&W^{(2)}(\boldsymbol{\rho}_{s1},\boldsymbol{\rho }_{i1},\boldsymbol{\rho }_{s2},\boldsymbol{\rho }_{i2},\theta _s,\theta _i,z )=Tr[\rho\\ &\hat{E}_{s1}^{( - )}(\theta _s,\boldsymbol{r}_{s1})\hat{E}_{i1}^{( - )}( \theta _i,\boldsymbol{r}_{i1} )\hat{E}_{i2}^{( + )}( \theta_i,\boldsymbol{r}_{i2}) \hat{E}_{s2}^{( +)}(\theta_s,\boldsymbol{r}_{s2})].
\label{EqA4}
\end{aligned}
\end{equation}

The term $ S^{( 2 )}(\cdot) $ in the Eq. (\ref{EqA3}) is denoted as the two-photon analogs of the spectral density function based the second-order coherence theory \cite{born1999Book}. In one of the cases where the down-converted photon pairs pass through the BS and are both detected by the detectors can be written as:
\begin{equation}
\begin{aligned}
&S^{(2)}(\boldsymbol{\rho }_{s1},\boldsymbol{\rho }_{i1},\theta _s,\theta _i,z ) =Tr[\rho\\ &\hat{E}_{s1}^{( - )}( \theta _s,\boldsymbol{r}_{s1} ) \hat{E}_{i1}^{( - )}( \theta _i,\boldsymbol{r}_{i1} ) \hat{E}_{i1}^{( + )}( \theta _i,\boldsymbol{r}_{i1} ) \hat{E}_{s1}^{( + )}( \theta _s,\boldsymbol{r}_{s1} )]
\label{EqA5}
\end{aligned}
\end{equation}

Furthermore, for the sake of further calculating the two-photon cross-spectral density function $ W^{( 2 )}( \boldsymbol{\rho }_{s1},\boldsymbol{\rho }_{i1},\boldsymbol{\rho }_{s2},\boldsymbol{\rho }_{i2},\theta _s,\theta _i,z ) $, the specific form of the positive frequency part of the signal and idler fields at the detection plane is written as:

\begin{equation}
\begin{aligned}
\hat{E}_{s1}^{( + )}( \theta _s,\boldsymbol{r}_{s1} )&=e^{ik_sz}\int{d\boldsymbol{q}_s\exp [ i( \boldsymbol{q}\cdot \boldsymbol{\rho }_{s1}-\frac{\boldsymbol{q}^2z}{2k_s} ) ] \hat{a}_{\theta _s,\boldsymbol{q}_s}}  \text{,  }
\\
\hat{E}_{i1}^{( + )}( \theta _i,\boldsymbol{r}_{i1} )&=e^{ik_iz}\int{d\boldsymbol{q}_i\exp [ i( \boldsymbol{q}\cdot \boldsymbol{\rho }_{i1}-\frac{\boldsymbol{q}^2z}{2k_i} ) ] \hat{a}_{\theta _i,\boldsymbol{q}_i} }\text{,   }
\label{EqA6}
\end{aligned}
\end{equation}
where $\hat{a}_{\theta _s,\boldsymbol{q}_s}$ denotes the annihilation operator for the mode with polarization state at angle $\theta _s$ with respect to the x-axis and transverse wave vector $\boldsymbol{q}_s$.
The differences of the trigonometric functions for the photons passing or reflecting through the BS, simultaneously, in the Eq. (\ref{EqA6}) comes from the type-II phase matching, which produces entangled states with mutually orthogonal polarizations.
Because of the narrow bandwidth filters inserted before the detectors, during the process of theoretical derivation, it is assumed that the down-converted photons are degenerated, which means the angular frequencies of the signal and idler satisfy the equation $ \omega _s=\omega _i={{\omega _p}\big/{2}} $. Meanwhile, when the phase matching condition is taken into account, one can conclude that the down-converted photon pairs can be considered paraxial, namely, $ k_s=k_i{{=k_p}\big/{2}} $. Substituting Eq. (\ref{EqA1}) and Eq. (\ref{EqA6}) into formula (\ref{EqA4}), the two-photon cross-spectral density function becomes:
\begin{equation}
\begin{aligned}
&W^{( 2 )}( \boldsymbol{\rho }_{s1},\boldsymbol{\rho }_{i1},\boldsymbol{\rho }_{s2},\boldsymbol{\rho }_{i2},\theta _s,\theta _i,z ) =\frac{| A |^2\sin 2\theta _s\sin 2\theta _i}{4}
\\
&\times \iiiint{d\boldsymbol{q}_sd\boldsymbol{q}_{s}^{'}d\boldsymbol{q}_id\boldsymbol{q}_{i}^{'}}< G( \boldsymbol{q}_s+\boldsymbol{q}_i ) G^*( \boldsymbol{q}_{s}^{'}+\boldsymbol{q}_{i}^{'} ) >
\\
&\times\exp [ i( \boldsymbol{q}_s\cdot \boldsymbol{\rho }_{s1}+\boldsymbol{q}_i\cdot \boldsymbol{\rho }_{i1}-\boldsymbol{q}_{s}^{'}\cdot \boldsymbol{\rho }_{s2}-\boldsymbol{q}_{i}^{'}\cdot \boldsymbol{\rho }_{i2} ) ]
\\
&\times \exp \{ -\frac{iz}{k_p}[ ( \boldsymbol{q}_{s}^{_2}+\boldsymbol{q}_{i}^{_2} ) -( \boldsymbol{q}_{s}^{'2}+\boldsymbol{q}_{i}^{'2} ) ] \},
\label{EqA7}
\end{aligned}
\end{equation}
where Eq. (\ref{EqA7}) is an integral of the angular correlation function of the pump field. It means that the properties of the pump field is transferred to the spatial coherence properties of the down-converted two-photon field.
Due to the absence of the theory that can perfectly describe the coherent characteristics of LED, we employ a general partially coherent pump beam of GSM for simulation and theory calculation. The cross-spectral density for a GSM pump beam, which is characterized by the transverse correlation length $ \ell _c $ at $ z=0 $ and beam waist width $ \sigma _0 $ at $ z=0 $ in momentum basis, can be expressed as \cite{mandel1995Book} :

\begin{equation}
\begin{aligned}
&<G( \boldsymbol{q}_s+\boldsymbol{q}_i ) G^*( \boldsymbol{q}_{s}^{'}+\boldsymbol{q}_{i}^{'} ) > =\frac{A_{p}^{2}\sigma _{0}^{2}\delta ^2}{4\pi}
\\
&\times \exp \{-\frac{\delta ^2(2\sigma _{0}^{2}+\ell _{c}^{2})}{4\ell _{c}^{2}}[ ( \boldsymbol{q}_s+\boldsymbol{q}_i ) ^2+( \boldsymbol{q}_{s}^{'}+\boldsymbol{q}_{i}^{'} ) ^2 ]\\
&+\frac{\delta ^2\sigma _{0}^{2}}{\ell _{c}^{2}}[ ( \boldsymbol{q}_s+\boldsymbol{q}_i ) \cdot ( \boldsymbol{q}_{s}^{'}+\boldsymbol{q}_{i}^{'} ) ] \},
\label{EqA8}
\end{aligned}
\end{equation}
where $ A_{p} $ is a constant. After substituting the Eq. (\ref{EqA8}) into Eq. (\ref{EqA7}), one can calculate the far-field expression of the two-photon cross-spectral density, $W^{(2)}(\boldsymbol{\rho }_{s1},\boldsymbol{\rho }_{i1},\boldsymbol{\rho }_{s2},\boldsymbol{\rho }_{i2},\theta _s,\theta _i,z)$ in the following form:

\begin{widetext}
\begin{equation}
\begin{aligned}
W^{( 2 )}( \boldsymbol{\rho }_{s1},\boldsymbol{\rho }_{i1},\boldsymbol{\rho }_{s2},\boldsymbol{\rho }_{i2},\theta _s,\theta _i,z ) &=| A |^2[ \frac{A_p\pi \sigma _0k_{p}^{2}\delta }{2z^2} ]^2\sin 2\theta _s\sin 2\theta _i\exp [ \frac{ik_p}{4z}( \boldsymbol{\rho }_{s1}^{2}+\boldsymbol{\rho }_{i1}^{2}-\boldsymbol{\rho }_{s2}^{2}-\boldsymbol{\rho }_{i2}^{2} ) ]
\\
&\times \exp \{ -\frac{\delta ^2k_{p}^{2}(\ell _{c}^{2}+2\sigma _{z}^{2})}{16z^2\ell _{c}^{2}}[ ( \boldsymbol{\rho }_{s1}+\boldsymbol{\rho }_{i1} ) ^2+( \boldsymbol{\rho }_{s2}+\boldsymbol{\rho }_{i2} ) ^2 ] \}
\\
&\times \exp \{-\frac{\delta ^2\sigma _{z}^{2}k_{p}^{2}}{4z^2\ell _{c}^{2}}[ ( \boldsymbol{\rho }_{s1}+\boldsymbol{\rho }_{i1} ) \cdot ( \boldsymbol{\rho }_{s2}+\boldsymbol{\rho }_{i2} ) ] \},
\label{EqA9}
\end{aligned}
\end{equation}
\end{widetext}
where $\sigma _{z}$ is the beam size at position z along the propagating direction, and $\sigma _z={{z\sqrt{\ell _{c}^{2}+4\sigma _{0}^{2}}}/{2k_p\sigma _0\ell _c}}$.
\begin{figure*}[t]
\centerline{\includegraphics[width=1.2\columnwidth]{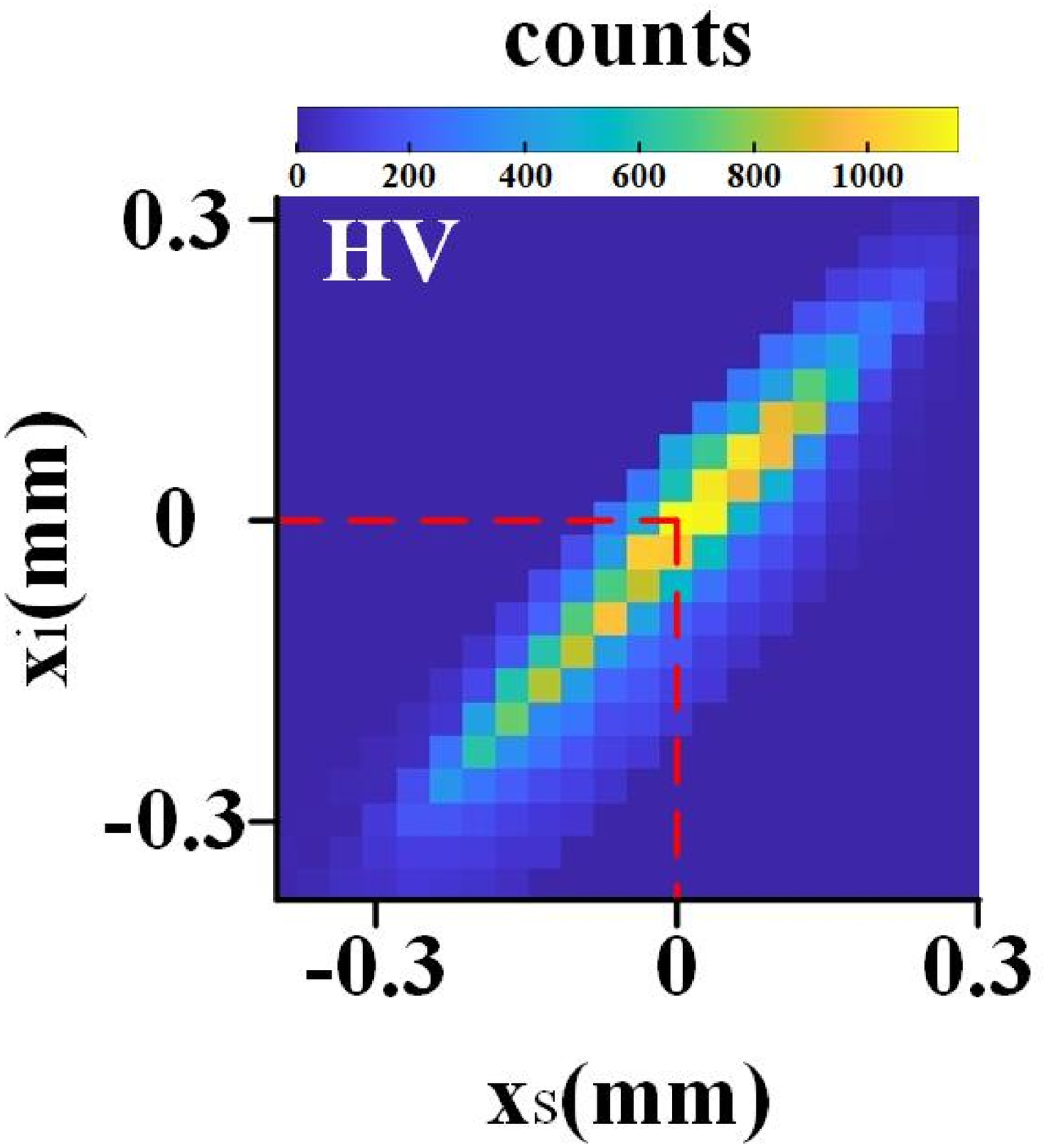}}
\caption{Experimental results of the measured counts when automatically scanning the two slits with the polarization angle $\theta _s=0,\theta _i=\pi/2$. Horizontal axes denote the position of the signal; vertical axes denote the position of the idler. The point corresponding to the red dotted line is the optimal position of the slits, which we used for the Bell measurement.}
\label{s2}
\end{figure*}
\begin{figure*}[t]
\centerline{\includegraphics[width=1\columnwidth]{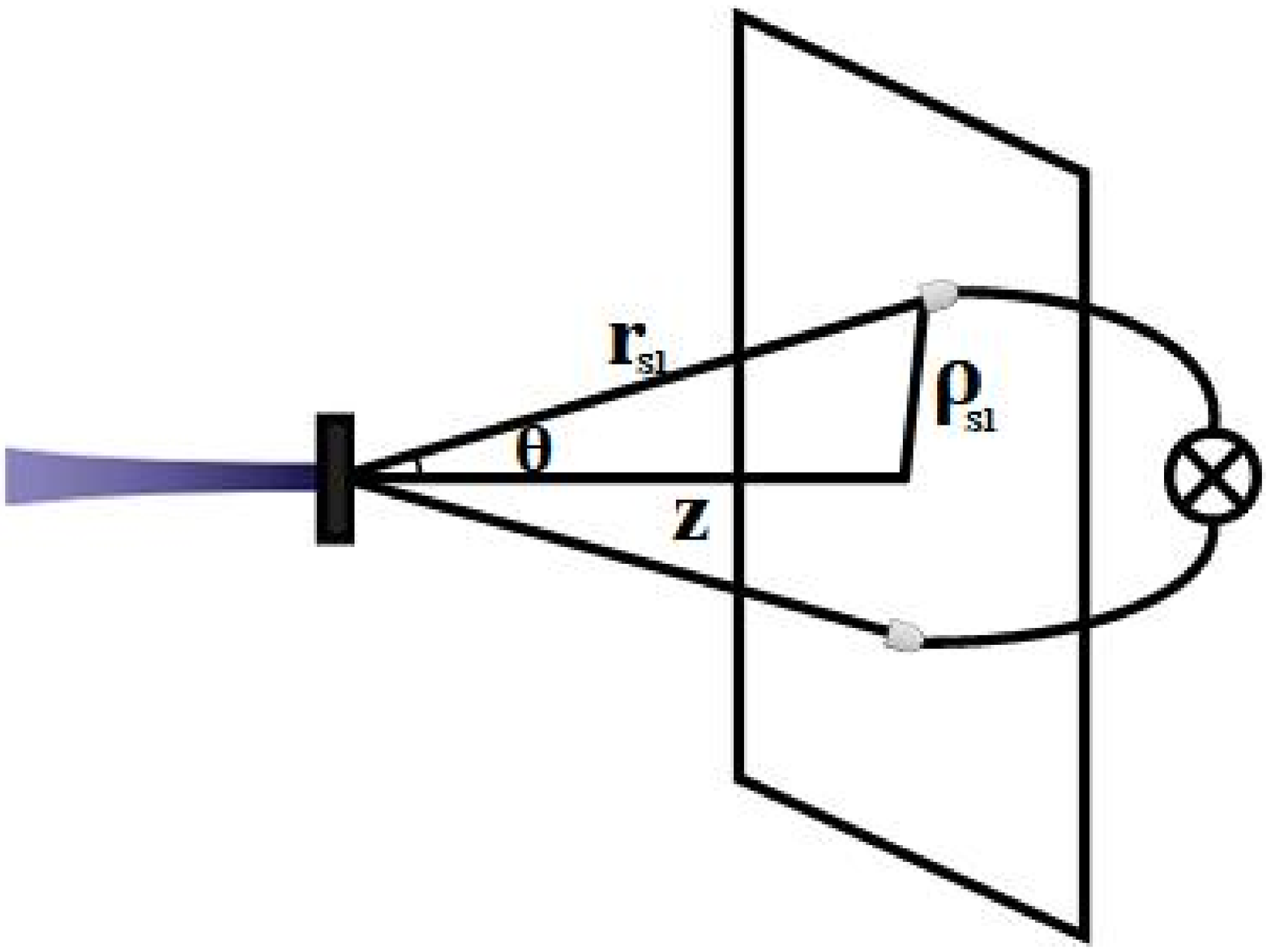}}
\caption{Schematic illustration of the relation between the propagation distance of the bi-photon and the detector's position.}
\label{s1}
\end{figure*}

Similarly, with the help of Eq. (\ref{EqA5}) together with (\ref{EqA9}) one can get the two types of the two-photon analogs of the spectral density functions in our experiment with the form as follows:
\begin{widetext}
\begin{equation}
\begin{aligned}
S^{( 2 )}( \boldsymbol{\rho }_{s1},\boldsymbol{\rho }_{i1},\theta _s,\theta _i,z ) &=| A |^2[\frac{2A_p\pi \sigma _{0}^{3}k_{p}^{2}\ell _{c}^{2}}{z^2\left( 4{\sigma _0}^2+\ell _{c}^{2} \right)} ] ^2\cos ^2\theta _s\sin ^2\theta _i\exp [ \frac{(\sigma _{z}^{2}-\ell _{c}^{2})\delta ^2k_{p}^{2}}{4z^2\ell _{c}^{2}}( \boldsymbol{\rho }_{s1}+\boldsymbol{\rho }_{i1} ) ^2 ]
\\
S^{( 2 )}( \boldsymbol{\rho }_{s2},\boldsymbol{\rho }_{i2},\theta _s,\theta _i,z ) &=| A |^2[ \frac{2A_p\pi \sigma _{0}^{3}k_{p}^{2}\ell _{c}^{2}}{z^2\left( 4{\sigma _0}^2+\ell _{c}^{2} \right)} ] ^2\sin ^2\theta _s\cos ^2\theta _i \exp [ \frac{(\sigma _{z}^{2}-\ell _{c}^{2})\delta ^2k_{p}^{2}}{4z^2\ell _{c}^{2}}( \boldsymbol{\rho }_{s2}+\boldsymbol{\rho }_{i2} ) ^2 ].
\label{EqA10}
\end{aligned}
\end{equation}
\end{widetext}

Firstly, we measure the position correlation of the bi-photons. The polarization angle is set as $\theta _s=0,\theta _i=\pi/2$, which is similar with our recent work that uses an LED to implement the position correlation detection of signal and idler photons  \cite{zhang2019influence}. The experimental results are shown in Fig. S\ref{s2}. Each point is obtained by accumulating 60 measurements. Each measurement is measured in 60s and averaged 10 times.

One can choose any one of the points as the red doted line labeled  to do the Bell measurement. In other words, we have a lot of the position to implement the polarization entanglement. However, it is noted that the count rate is too small to obtain for the edge position. So, in our experiment, we choose the center one, which is the highest count rate for the following experiment. By confirming the position of the single slits, the transverse position of the photon pairs can be assumed as $\boldsymbol{\rho }_{s1}+\boldsymbol{\rho }_{i1}=0$, and similarly, $\boldsymbol{\rho }_{s2}+\boldsymbol{\rho }_{i2}=0$.
Besides, in our experiment, the distance from the nonlinear crystal to the detectors is much longer than the beam size, so we have  $\boldsymbol{r}_{s1}\gg \boldsymbol{\rho }_{s1}$ and $\theta\approx0$. The relation between the propagation distance of the bi-photon and the detector's position can be found in Fig. S\ref{s1}. Then we perform a Taylor expansion and make the approximation as $ \boldsymbol{r}_{s1}\approx z+{{ \boldsymbol{\rho}_{s1}^{2}} \big/ {2z}}$. So, the first exponential term of the two-photon cross-spectral density of Eq. (\ref{EqA9}) can be written as $\exp \left[ \frac{i\boldsymbol{k}_p}{2}\left( \boldsymbol{r}_{s1}^{2}+\boldsymbol{r}_{i1}^{2}-\boldsymbol{r}_{s2}^{2}-\boldsymbol{r}_{i2}^{2} \right) \right] $. Furthermore, the parameter $\boldsymbol{r}_{s}$, which represents the position of the detector $ D_s $, is defined as $\boldsymbol{r}_s={{\left( \boldsymbol{r}_{s1}-\boldsymbol{r}_{s2} \right)}/{2}}$. Similarly, the parameter $\boldsymbol{r}_{i}$ can be also defined as $\boldsymbol{r}_i={{\left( \boldsymbol{r}_{i1}-\boldsymbol{r}_{i2} \right)}/{2}}$. Then, the term above ends up being written as $\exp \left[ i\boldsymbol{k}_p\left( \boldsymbol{r}_s+\boldsymbol{r}_i \right) \right] $. After rewriting the formula (\ref{EqA9}) and (\ref{EqA10}) with the parameters defined above, substituting the two formulas into formula (\ref{EqA3}), one can obtain the precise function of $R_{si}$ as:
\begin{widetext}
\begin{equation}
\begin{aligned}
R_{si}(\theta_s,\theta _i ,\boldsymbol{r}_s,\boldsymbol{r}_i)& =\Bigg|| A |^2e^{ik_p(\boldsymbol{r}_s+\boldsymbol{r}_i )}[ \frac{A_p\pi \ell _C{\sigma _0}^2}{z\sqrt{4{\sigma _0}^2+\ell _{c}^{2}}}] ^2\\
&\times[C_1C_2\cdot \frac{\pi z^2}{2k_{p}^{2}\sigma _z\delta}\sin 2\theta _s\sin 2\theta_i+\sqrt{\frac{2\pi}{k_{p}^{2}\delta ^2}}(C_{1}^{2}\cos ^2\theta _s\sin ^2\theta _i+C_{2}^{2}\sin ^2\theta _s\cos ^2\theta_i)]\Bigg|.
\label{EqA11}
\end{aligned}
\end{equation}
\end{widetext}

Based on the Eq. (\ref{EqA11}), one can actually evaluate the quality of the polarization entanglement, generated by the SPDC process pumped by LED. We consider to measure the Clauser, Horne, Shimony, and Holt (CHSH) form of Bell's inequality \cite{Clauser1969PhysRevLett}. For ease of calculation, the CHSH expectation E function is defined by:
\begin{equation}
\begin{aligned}
E( \theta _s,\theta _i ) =\frac{R( \theta _s,\theta _i ) +R( \theta _{s}^{_{\bot}},\theta _{i}^{_{\bot}} ) -R( \theta _{s}^{_{\bot}},\theta _i ) -R( \theta _s,\theta _{i}^{_{\bot}} )}{R( \theta _s,\theta _i ) +R( \theta _{s}^{_{\bot}},\theta _{i}^{_{\bot}} ) +R( \theta _{s}^{_{\bot}},\theta _i ) +R( \theta _s,\theta _{i}^{_{\bot}} )},
\label{EqA12}
\end{aligned}
\end{equation}
where the term $R(\theta_s,\theta_i )$ represents the coincidence count rate of the down converted state when the analyzers are placed in the path of the signal and idler making respective angles $\theta_s$ and $\theta_i$ with x axis, respectively, while the parameters $\theta _{s}^{_{\bot}}$ and $\theta_{i}^{_{\bot}}$ are the corresponding orthogonal angles. Moreover, $E(\theta_s,\theta_i )$ represents the quantum correlation between the photons with their polarization in the respective basis. After that, the parameter S is composed for two values of $\theta_s$ and two values of $\theta_i$, which is usually defined as:
\begin{equation}
\begin{aligned}
S=|E(\theta_s,\theta_i)-E(\theta_{s}^{'},\theta _i ) +E(\theta_s,\theta_{i}^{'}) +E(\theta _{s}^{'},\theta_{i}^{'})|.
\label{EqA13}
\end{aligned}
\end{equation}
Classical and hidden variable theories predict $S\leqslant2$, while quantum mechanics permits $ 2<S\leqslant 2\sqrt{2}$, with equality occurring for maximal polarization entanglement, namely, the polarization Bell state.
In our case of measuring polarization entanglement, the down-converted photons in the transmitted path and in the reflected path of the 50:50 beam splitter are separately analyzed with a HWP and a PBS. Also, the pump beam wavelength $\lambda _p$ is 405 nm. The transverse coherent length $\ell _c$ is estimated to be 11 $\mu m$ based on our previous work \cite{zhang2019influence}, and beam waist $\sigma _0$ is measured few times and averaged approximately as 5 mm. And in our experiment we adopt $\theta _s=0,{{\pi}\big/{8}}$ and $\theta _i={{\pi}\big/{16}},{{3\pi}\big/{16}}$. Bringing these specific values into the function of S, which can obtain the value as 2.61. It is noted that we have obtained that the S is about 2.3 in experiment. In our measurement setup, the slit as well as the detection system is not a strictly point detector as the theoretical assumption. Therefore, the collection diameter determines the number of the collected modes. The purity of the collected modes lowers the value of  concurrence and Bell parameter. We expect the value can be improved with smaller collection setups but will loss the whole system detection efficiency.


%